\documentstyle[12pt]{article}
\begin{document}
\begin{center} {\Large {\bf Cluster Statistics  of BTW Automata}} \end{center}
\begin{center} {\it Ajanta Bhowal Acharyya\\
Department of Physics, Lady Brabourne College,\\ 
P-1/2 Surahwardy Avenue, Calcutta-700017, India\\
ajanta.bhowal@gmail.com}\end{center}

\vskip 1cm

\noindent The cluster statistics of BTW automata 
in the SOC states are obtained
by extensive computer simulation. Various moments of the 
clusters are calculated and few results
are compared with earlier available numerical estimates 
and exact results. Reasonably good agreement
is observed. An extended statistical analysis has been made.

\vskip 1cm



\vskip 2cm

\noindent {\bf I. Introduction:}

Bak, Tang and Wiesenfeld (BTW) [1] have recently (1987) introduced the
concept of Self Organised Criticality  to describe the appearance of 
long range spatio-temporal correlations observed in extended,
dissipative dynamical systems in nature. 
This phenomena of SOC is characterised by spontaneous evolution
into a steady state which shows long range temporal and spatial
correlation.  The simple lattice automata model of sandpile which shows
this SOC behaviour was studied by BTW[1].  Substantial developments have
been made on the study of BTW model.  Several properties of this critical
state, e.g., entropy, height correlation, height probabilities
etc  have been  calculated analytically. But the critical exponents
have not been calculated analytically. Extensive numerical efforts have been 
performed to estimate various exponents. 
Recently the studies regarding this model have been reviewed by Dhar[2].

Here, in this paper, the statistics of the clusters, formed by the sites having 
particular value of the automaton, has been studied, by computer simulation.
We have  calculated the various statistical quantities e.g., average size
of clusters, total numbers of clusters, maximum size of the clusters etc. 
in the SOC state and found the distribution function of the clusters.
Related to the statistics of the clusters of the automation values,
Manna has estimated the fraction of sites, having automation values, 0,1,2 and
3, present in the SOC state[3,4]. Majumder and Dhar calculated exactly the
fraction of sites having minimum value of automation in the SOC state[5].

The paper is organised as follows: In section II the model and the simulation
technique have been described, section III contains the numerical results 
obtained here, and the paper ends with a concluding section (section IV).
\vspace { 1 cm }

\noindent {\bf II. The Model and Simulation:}

BTW lattice automata model [1]  evolves to a stationary
state in a self-organised (having no tunnable parameter) way. This state has
no scale of length and time, hence called critical. Altogether the state is
called self-organised critical state.  The description of the lattice automata 
model is as follows : At each site of this lattice, a variable
(automaton) $z(i,j)$ is associated which can take positive integer values.
Starting from the initial condition (at every site $z(i,j)$ = 0), the value of $z(i,j)$
is increased (so called addition of one 'sand' particle) at randomly chosen
site $(i,j)$ of the lattice in steps of unity as,
$$z(i,j) = z(i,j) + 1.$$
\noindent When the value of $z$ at any site reaches a maximum $z_m$, its
value decreases by four units (i.e., it topples) and each of the four nearest
neighbours gets one unit of $z$ (maintaining local conservation) as follows:
$$z(i,j) = z(i,j) - 4$$
$$z(i\pm 1, j\pm 1) = z(i\pm 1, j\pm 1) + 1 \eqno(1)$$
\noindent for $z(i,j) \geq z_m$. Each boundary site is attatched with an
 additional site which acts as a sink (i.e., open boundary condition is used ). 
In this simulation, the value of $z_m$ = 4. 

It has been observed that, as the time goes on the 
average value ($\bar z$) of $z(i,j)$, over the space, increases and ultimately
reaches a steady value ($\bar z_c$) characterising the SOC state. 
At any time (i.e., a particular value of $\bar z $), after the addition
 of one such 'sand' particle, when the system becomes stable (i.e., all the
 sites have $z(i,j)<4$), the lattice contains
the sites having the automaton values 0, 1, 2 and 3 only. These sites (having a
particular value of $z(i,j)$) form clusters connected via the 
nearest neighbours. These clusters have some size distribution described by
the function $\rho_k (s_k)$, where $\rho_k (s_k)$ denotes the
number  of clusters (formed via nearest neighbour connection of
sites having their automaton value $k$) of size $s_k$. 
From this distribution $\rho_k(s_k)$, 
the various statistical moments of such clusters have
been calculated here and studied as a function of $\bar z$.
The various statistical quantities (moments), 
for a particular class of cluster,
are defined as follows:

\begin{itemize}
\item{} Total fraction of sites ($F_k(\bar z)$) having value $k$ at a 
particular value of $\bar z$.
$$F_k(\bar z) = {1 \over L^2} \left(\sum s_k \rho_k (s_k)\right)$$
\item{} Average size ($S_k(\bar z)$) of clusters formed by the sites 
of automation value $k$. If $\rho_k (s_k)$ is the number of clusters (formed
by the sites having their automation value $z$) of size $s_k$. 
$$S_k (\bar z) =  {{\sum s_k \rho_k (s_k)} 
\over {\sum \rho_k (s_k)}} $$
\item{} Total number of clusters ($N_k(\bar z)$) formed by the sites having
their automaton value $z(i,j) = k$.
$$N_k (\bar z) = \sum \rho_k (s_k)$$
\item{} Largest size ($s^{max}_k(\bar z)$) of cluster formed by the 
sites of their automation value $k$.
$$L_k (\bar z)  =  Max [s_k]$$
\end{itemize}
\vspace { 1 cm }

\noindent {\bf III. Results:}

We have calculated the cluster size distribution, $\rho_k(s_k)$, in the SOC
state for different automaton values ($K$ = 0, 1, 2 and 3). We did it by using
the program [6] to calculate the cluster size distribution used in standard
percolation problem [7]. From the distribution all the statistical quantities
(or moments) i.e., $F_k$, $S_k$, $N_k$ and $L_k$ are calculated in the SOC state.
All these data has been obtained for L=400, by averaging over    
20 different random samples. 
The values of $F_k$ have been estimated earlier [3,4,5] and our values
agree well with the previous estimates (see Table I).

Fig.1 shows the semilog plot of the normalised cluster size distribution,
($n_s=\rho_k(s_k)/L^2$, is the number of clusters of size s per lattice site),
for the automation value, k=3, and for L=400. The distribution function
fits with the curve 
$$n_s=.000004*s^{-1.5}*exp(-.045*s).$$

To compare the cluster size distribution with the known result of percolation, 
we have plotted here the normalised cluster size distribution, $n_s = \rho_k(s_k)/L^2$,
instead of $\rho_k(s_k)$. We see that this distribution function (for $k$ = 3)
has the same functional form as that of the percolation [7] (below $p_c$) apart
from the exponent $\theta$ of the pre-exponential factor, which may be the due to
the fact that the sites are not randomly filled and they are correlated [9]. 
Another interesting point to note
here is that the clusters (for $k$ = 0, 1, 2, 3) formed by automaton values
are not system spanning (see the first column of table-I), although it can give
a system spanning (scale invariant) avalanche in SOC state.
It is to be mentioned that the other two distribution functions of the clusters,
for automation value k=2 and 1, have also been observed. These two distribution
functions are also of the same nature as that of k=3. Since the maximum size
of the cluster decreases as the automaton value decreases, the fluctuation
in the distribution function for large clusters increases and hence these
results have not shown explicitly.

The basic features of this simulational studies may be described
briefly as follows: In the self organised critical state the BTW
automata will assume the values 0, 1, 2 and 3. These automata will
form the clusters. The statistics of such clusters are studied here.
The distribution of the sizes of clusters, average size, maximum size
etc. are studied by numerical simulation. An algebraic form of the
cluster size distribution is proposed. Interestingly, it was observed
that the clusters are not system spanning in the SOC state.

\vskip 2cm

\noindent {\bf IV. Concluding Remarks:}

The statistical aspects of BTW cellular automata is investigated by computer silmulation. In the
SOC state the automaton can take the values 0, 1, 2 and 3. The cluster formaed by the fixed value
of automaton (say 3) is identified and measured its size. The clusters are defined via nearest
neighbour connections. The statistical distribution of these clusters are calculated and from
this distribution various moments are calculated in SOC state. Few results (for example the fraction
of sites having a fixed value) are compared 
with the earlier numerical estimates (Ref.3-4) and exact
calculations (Ref.5). The values of $S_k^{SOC}$ and 
$L_k^{SOC}$ for k=0 listed in table I are
exactly known from Dhar's [2] burning algorithm.
This study is a generalisation of earlier study (which estimates only fraction of sites having a fixed
value) extending the calculation of entire statistics of BTW automata. 
It would be interesting to study the dynamical evolution (towards SOC state) of the cluster, its formation
nucleation, coalescence etc.
\newpage

\begin{center} {\bf {Table I}} \end{center}
\begin{center}
\begin{tabular}{|c|c|c|c|c|}
\hline

$k$ & $F_k^{SOC}$ & $S_k^{SOC}$ & $N_k^{SOC}$ & $L_k^{SOC}$ \\

\hline

0 & 0.0741 & 1.000 & 11856.2 & 1.00 \\
 & 0.0730$^a$ &  &  &  \\
 & 0.0736$^b$ &  &  & \\
 & 0.0736$^c$ &  &  &  \\
 & 0.07363$^d$ &  &  &  \\

\hline

1 & 0.1746 & 1.429 & 19552.4 & 10.85 \\
 & 0.1740$^a$ &  &  &  \\
 & 0.1740$^b$ &  &  &  \\
 & 0.1739$^d$ &  &  &  \\

\hline

2 & 0.3069 & 2.376 & 20670.2 & 37.90 \\
 & 0.307$^a$ &  &  &  \\
 & 0.3062$^b$ &  &  &  \\
 & 0.3063$^d$ &  &  &  \\

\hline

3 & 0.444 & 4.326 & 16430.5 & 109.84 \\
 & 0.446$^a$ &  &  &  \\
 & 0.4462$^b$ &  &  &  \\
 & 0.4461$^d$ &  &  &  \\
\hline
\end{tabular}
\end{center}
\begin{flushleft}
{\bf $^a$S. S. Manna, J. Stat. Phys., 59 (1990) 509} \\
{\bf $^b$P. Grassberger \& S. S.Manna J. de. Physique, 51 (1990) 1077} \\
{\bf $^c$S. N. Majumder \& D. Dhar, J. Phys. A:Math. Gen {\bf 24} 
(1990) L357} \\
{\bf $^d$ V. B. Priezzhev, J. Stat. Phys. 74 (1994) 955}\\
\end{flushleft}

\newpage
\setlength{\unitlength}{0.240900pt}
\ifx\plotpoint\undefined\newsavebox{\plotpoint}\fi
\sbox{\plotpoint}{\rule[-0.200pt]{0.400pt}{0.400pt}}%
\begin{picture}(1500,900)(0,0)
\font\gnuplot=cmr10 at 10pt
\gnuplot
\sbox{\plotpoint}{\rule[-0.200pt]{0.400pt}{0.400pt}}%
\put(221.0,122.0){\rule[-0.200pt]{4.818pt}{0.400pt}}
\put(201,122){\makebox(0,0)[r]{1e-12}}
\put(1420.0,122.0){\rule[-0.200pt]{4.818pt}{0.400pt}}
\put(221.0,154.0){\rule[-0.200pt]{2.409pt}{0.400pt}}
\put(1430.0,154.0){\rule[-0.200pt]{2.409pt}{0.400pt}}
\put(221.0,196.0){\rule[-0.200pt]{2.409pt}{0.400pt}}
\put(1430.0,196.0){\rule[-0.200pt]{2.409pt}{0.400pt}}
\put(221.0,217.0){\rule[-0.200pt]{2.409pt}{0.400pt}}
\put(1430.0,217.0){\rule[-0.200pt]{2.409pt}{0.400pt}}
\put(221.0,227.0){\rule[-0.200pt]{4.818pt}{0.400pt}}
\put(201,227){\makebox(0,0)[r]{1e-11}}
\put(1420.0,227.0){\rule[-0.200pt]{4.818pt}{0.400pt}}
\put(221.0,259.0){\rule[-0.200pt]{2.409pt}{0.400pt}}
\put(1430.0,259.0){\rule[-0.200pt]{2.409pt}{0.400pt}}
\put(221.0,301.0){\rule[-0.200pt]{2.409pt}{0.400pt}}
\put(1430.0,301.0){\rule[-0.200pt]{2.409pt}{0.400pt}}
\put(221.0,322.0){\rule[-0.200pt]{2.409pt}{0.400pt}}
\put(1430.0,322.0){\rule[-0.200pt]{2.409pt}{0.400pt}}
\put(221.0,333.0){\rule[-0.200pt]{4.818pt}{0.400pt}}
\put(201,333){\makebox(0,0)[r]{1e-10}}
\put(1420.0,333.0){\rule[-0.200pt]{4.818pt}{0.400pt}}
\put(221.0,364.0){\rule[-0.200pt]{2.409pt}{0.400pt}}
\put(1430.0,364.0){\rule[-0.200pt]{2.409pt}{0.400pt}}
\put(221.0,406.0){\rule[-0.200pt]{2.409pt}{0.400pt}}
\put(1430.0,406.0){\rule[-0.200pt]{2.409pt}{0.400pt}}
\put(221.0,428.0){\rule[-0.200pt]{2.409pt}{0.400pt}}
\put(1430.0,428.0){\rule[-0.200pt]{2.409pt}{0.400pt}}
\put(221.0,438.0){\rule[-0.200pt]{4.818pt}{0.400pt}}
\put(201,438){\makebox(0,0)[r]{1e-09}}
\put(1420.0,438.0){\rule[-0.200pt]{4.818pt}{0.400pt}}
\put(221.0,470.0){\rule[-0.200pt]{2.409pt}{0.400pt}}
\put(1430.0,470.0){\rule[-0.200pt]{2.409pt}{0.400pt}}
\put(221.0,511.0){\rule[-0.200pt]{2.409pt}{0.400pt}}
\put(1430.0,511.0){\rule[-0.200pt]{2.409pt}{0.400pt}}
\put(221.0,533.0){\rule[-0.200pt]{2.409pt}{0.400pt}}
\put(1430.0,533.0){\rule[-0.200pt]{2.409pt}{0.400pt}}
\put(221.0,543.0){\rule[-0.200pt]{4.818pt}{0.400pt}}
\put(201,543){\makebox(0,0)[r]{1e-08}}
\put(1420.0,543.0){\rule[-0.200pt]{4.818pt}{0.400pt}}
\put(221.0,575.0){\rule[-0.200pt]{2.409pt}{0.400pt}}
\put(1430.0,575.0){\rule[-0.200pt]{2.409pt}{0.400pt}}
\put(221.0,617.0){\rule[-0.200pt]{2.409pt}{0.400pt}}
\put(1430.0,617.0){\rule[-0.200pt]{2.409pt}{0.400pt}}
\put(221.0,638.0){\rule[-0.200pt]{2.409pt}{0.400pt}}
\put(1430.0,638.0){\rule[-0.200pt]{2.409pt}{0.400pt}}
\put(221.0,648.0){\rule[-0.200pt]{4.818pt}{0.400pt}}
\put(201,648){\makebox(0,0)[r]{1e-07}}
\put(1420.0,648.0){\rule[-0.200pt]{4.818pt}{0.400pt}}
\put(221.0,680.0){\rule[-0.200pt]{2.409pt}{0.400pt}}
\put(1430.0,680.0){\rule[-0.200pt]{2.409pt}{0.400pt}}
\put(221.0,722.0){\rule[-0.200pt]{2.409pt}{0.400pt}}
\put(1430.0,722.0){\rule[-0.200pt]{2.409pt}{0.400pt}}
\put(221.0,744.0){\rule[-0.200pt]{2.409pt}{0.400pt}}
\put(1430.0,744.0){\rule[-0.200pt]{2.409pt}{0.400pt}}
\put(221.0,754.0){\rule[-0.200pt]{4.818pt}{0.400pt}}
\put(201,754){\makebox(0,0)[r]{1e-06}}
\put(1420.0,754.0){\rule[-0.200pt]{4.818pt}{0.400pt}}
\put(221.0,785.0){\rule[-0.200pt]{2.409pt}{0.400pt}}
\put(1430.0,785.0){\rule[-0.200pt]{2.409pt}{0.400pt}}
\put(221.0,827.0){\rule[-0.200pt]{2.409pt}{0.400pt}}
\put(1430.0,827.0){\rule[-0.200pt]{2.409pt}{0.400pt}}
\put(221.0,849.0){\rule[-0.200pt]{2.409pt}{0.400pt}}
\put(1430.0,849.0){\rule[-0.200pt]{2.409pt}{0.400pt}}
\put(221.0,859.0){\rule[-0.200pt]{4.818pt}{0.400pt}}
\put(201,859){\makebox(0,0)[r]{1e-05}}
\put(1420.0,859.0){\rule[-0.200pt]{4.818pt}{0.400pt}}
\put(221.0,122.0){\rule[-0.200pt]{0.400pt}{4.818pt}}
\put(221,81){\makebox(0,0){0}}
\put(221.0,839.0){\rule[-0.200pt]{0.400pt}{4.818pt}}
\put(356.0,122.0){\rule[-0.200pt]{0.400pt}{4.818pt}}
\put(356,81){\makebox(0,0){20}}
\put(356.0,839.0){\rule[-0.200pt]{0.400pt}{4.818pt}}
\put(492.0,122.0){\rule[-0.200pt]{0.400pt}{4.818pt}}
\put(492,81){\makebox(0,0){40}}
\put(492.0,839.0){\rule[-0.200pt]{0.400pt}{4.818pt}}
\put(627.0,122.0){\rule[-0.200pt]{0.400pt}{4.818pt}}
\put(627,81){\makebox(0,0){60}}
\put(627.0,839.0){\rule[-0.200pt]{0.400pt}{4.818pt}}
\put(763.0,122.0){\rule[-0.200pt]{0.400pt}{4.818pt}}
\put(763,81){\makebox(0,0){80}}
\put(763.0,839.0){\rule[-0.200pt]{0.400pt}{4.818pt}}
\put(898.0,122.0){\rule[-0.200pt]{0.400pt}{4.818pt}}
\put(898,81){\makebox(0,0){100}}
\put(898.0,839.0){\rule[-0.200pt]{0.400pt}{4.818pt}}
\put(1034.0,122.0){\rule[-0.200pt]{0.400pt}{4.818pt}}
\put(1034,81){\makebox(0,0){120}}
\put(1034.0,839.0){\rule[-0.200pt]{0.400pt}{4.818pt}}
\put(1169.0,122.0){\rule[-0.200pt]{0.400pt}{4.818pt}}
\put(1169,81){\makebox(0,0){140}}
\put(1169.0,839.0){\rule[-0.200pt]{0.400pt}{4.818pt}}
\put(1305.0,122.0){\rule[-0.200pt]{0.400pt}{4.818pt}}
\put(1305,81){\makebox(0,0){160}}
\put(1305.0,839.0){\rule[-0.200pt]{0.400pt}{4.818pt}}
\put(1440.0,122.0){\rule[-0.200pt]{0.400pt}{4.818pt}}
\put(1440,81){\makebox(0,0){180}}
\put(1440.0,839.0){\rule[-0.200pt]{0.400pt}{4.818pt}}
\put(221.0,122.0){\rule[-0.200pt]{293.657pt}{0.400pt}}
\put(1440.0,122.0){\rule[-0.200pt]{0.400pt}{177.543pt}}
\put(221.0,859.0){\rule[-0.200pt]{293.657pt}{0.400pt}}
\put(41,490){\makebox(0,0){$n_s$}}
\put(830,40){\makebox(0,0){$s$}}
\put(221.0,122.0){\rule[-0.200pt]{0.400pt}{177.543pt}}
\put(228,804){\raisebox{-.8pt}{\makebox(0,0){$\bullet$}}}
\put(235,744){\raisebox{-.8pt}{\makebox(0,0){$\bullet$}}}
\put(241,724){\raisebox{-.8pt}{\makebox(0,0){$\bullet$}}}
\put(248,707){\raisebox{-.8pt}{\makebox(0,0){$\bullet$}}}
\put(255,694){\raisebox{-.8pt}{\makebox(0,0){$\bullet$}}}
\put(262,682){\raisebox{-.8pt}{\makebox(0,0){$\bullet$}}}
\put(268,669){\raisebox{-.8pt}{\makebox(0,0){$\bullet$}}}
\put(275,659){\raisebox{-.8pt}{\makebox(0,0){$\bullet$}}}
\put(282,651){\raisebox{-.8pt}{\makebox(0,0){$\bullet$}}}
\put(289,643){\raisebox{-.8pt}{\makebox(0,0){$\bullet$}}}
\put(295,635){\raisebox{-.8pt}{\makebox(0,0){$\bullet$}}}
\put(302,626){\raisebox{-.8pt}{\makebox(0,0){$\bullet$}}}
\put(309,621){\raisebox{-.8pt}{\makebox(0,0){$\bullet$}}}
\put(316,614){\raisebox{-.8pt}{\makebox(0,0){$\bullet$}}}
\put(323,606){\raisebox{-.8pt}{\makebox(0,0){$\bullet$}}}
\put(329,603){\raisebox{-.8pt}{\makebox(0,0){$\bullet$}}}
\put(336,598){\raisebox{-.8pt}{\makebox(0,0){$\bullet$}}}
\put(343,591){\raisebox{-.8pt}{\makebox(0,0){$\bullet$}}}
\put(350,586){\raisebox{-.8pt}{\makebox(0,0){$\bullet$}}}
\put(356,583){\raisebox{-.8pt}{\makebox(0,0){$\bullet$}}}
\put(363,574){\raisebox{-.8pt}{\makebox(0,0){$\bullet$}}}
\put(370,566){\raisebox{-.8pt}{\makebox(0,0){$\bullet$}}}
\put(377,565){\raisebox{-.8pt}{\makebox(0,0){$\bullet$}}}
\put(384,559){\raisebox{-.8pt}{\makebox(0,0){$\bullet$}}}
\put(390,556){\raisebox{-.8pt}{\makebox(0,0){$\bullet$}}}
\put(397,551){\raisebox{-.8pt}{\makebox(0,0){$\bullet$}}}
\put(404,545){\raisebox{-.8pt}{\makebox(0,0){$\bullet$}}}
\put(411,544){\raisebox{-.8pt}{\makebox(0,0){$\bullet$}}}
\put(417,537){\raisebox{-.8pt}{\makebox(0,0){$\bullet$}}}
\put(424,533){\raisebox{-.8pt}{\makebox(0,0){$\bullet$}}}
\put(431,531){\raisebox{-.8pt}{\makebox(0,0){$\bullet$}}}
\put(438,526){\raisebox{-.8pt}{\makebox(0,0){$\bullet$}}}
\put(444,517){\raisebox{-.8pt}{\makebox(0,0){$\bullet$}}}
\put(451,509){\raisebox{-.8pt}{\makebox(0,0){$\bullet$}}}
\put(458,510){\raisebox{-.8pt}{\makebox(0,0){$\bullet$}}}
\put(465,502){\raisebox{-.8pt}{\makebox(0,0){$\bullet$}}}
\put(472,496){\raisebox{-.8pt}{\makebox(0,0){$\bullet$}}}
\put(478,504){\raisebox{-.8pt}{\makebox(0,0){$\bullet$}}}
\put(485,492){\raisebox{-.8pt}{\makebox(0,0){$\bullet$}}}
\put(492,488){\raisebox{-.8pt}{\makebox(0,0){$\bullet$}}}
\put(499,489){\raisebox{-.8pt}{\makebox(0,0){$\bullet$}}}
\put(505,480){\raisebox{-.8pt}{\makebox(0,0){$\bullet$}}}
\put(512,477){\raisebox{-.8pt}{\makebox(0,0){$\bullet$}}}
\put(519,472){\raisebox{-.8pt}{\makebox(0,0){$\bullet$}}}
\put(526,472){\raisebox{-.8pt}{\makebox(0,0){$\bullet$}}}
\put(533,462){\raisebox{-.8pt}{\makebox(0,0){$\bullet$}}}
\put(539,456){\raisebox{-.8pt}{\makebox(0,0){$\bullet$}}}
\put(546,453){\raisebox{-.8pt}{\makebox(0,0){$\bullet$}}}
\put(553,455){\raisebox{-.8pt}{\makebox(0,0){$\bullet$}}}
\put(560,452){\raisebox{-.8pt}{\makebox(0,0){$\bullet$}}}
\put(566,449){\raisebox{-.8pt}{\makebox(0,0){$\bullet$}}}
\put(573,453){\raisebox{-.8pt}{\makebox(0,0){$\bullet$}}}
\put(580,431){\raisebox{-.8pt}{\makebox(0,0){$\bullet$}}}
\put(587,438){\raisebox{-.8pt}{\makebox(0,0){$\bullet$}}}
\put(593,419){\raisebox{-.8pt}{\makebox(0,0){$\bullet$}}}
\put(600,428){\raisebox{-.8pt}{\makebox(0,0){$\bullet$}}}
\put(607,428){\raisebox{-.8pt}{\makebox(0,0){$\bullet$}}}
\put(614,424){\raisebox{-.8pt}{\makebox(0,0){$\bullet$}}}
\put(621,404){\raisebox{-.8pt}{\makebox(0,0){$\bullet$}}}
\put(627,421){\raisebox{-.8pt}{\makebox(0,0){$\bullet$}}}
\put(634,396){\raisebox{-.8pt}{\makebox(0,0){$\bullet$}}}
\put(641,402){\raisebox{-.8pt}{\makebox(0,0){$\bullet$}}}
\put(648,412){\raisebox{-.8pt}{\makebox(0,0){$\bullet$}}}
\put(654,400){\raisebox{-.8pt}{\makebox(0,0){$\bullet$}}}
\put(661,398){\raisebox{-.8pt}{\makebox(0,0){$\bullet$}}}
\put(668,400){\raisebox{-.8pt}{\makebox(0,0){$\bullet$}}}
\put(675,386){\raisebox{-.8pt}{\makebox(0,0){$\bullet$}}}
\put(682,394){\raisebox{-.8pt}{\makebox(0,0){$\bullet$}}}
\put(688,394){\raisebox{-.8pt}{\makebox(0,0){$\bullet$}}}
\put(695,383){\raisebox{-.8pt}{\makebox(0,0){$\bullet$}}}
\put(702,377){\raisebox{-.8pt}{\makebox(0,0){$\bullet$}}}
\put(709,366){\raisebox{-.8pt}{\makebox(0,0){$\bullet$}}}
\put(715,362){\raisebox{-.8pt}{\makebox(0,0){$\bullet$}}}
\put(722,352){\raisebox{-.8pt}{\makebox(0,0){$\bullet$}}}
\put(729,366){\raisebox{-.8pt}{\makebox(0,0){$\bullet$}}}
\put(736,366){\raisebox{-.8pt}{\makebox(0,0){$\bullet$}}}
\put(742,362){\raisebox{-.8pt}{\makebox(0,0){$\bullet$}}}
\put(749,339){\raisebox{-.8pt}{\makebox(0,0){$\bullet$}}}
\put(756,339){\raisebox{-.8pt}{\makebox(0,0){$\bullet$}}}
\put(763,288){\raisebox{-.8pt}{\makebox(0,0){$\bullet$}}}
\put(770,288){\raisebox{-.8pt}{\makebox(0,0){$\bullet$}}}
\put(776,320){\raisebox{-.8pt}{\makebox(0,0){$\bullet$}}}
\put(783,307){\raisebox{-.8pt}{\makebox(0,0){$\bullet$}}}
\put(790,339){\raisebox{-.8pt}{\makebox(0,0){$\bullet$}}}
\put(797,330){\raisebox{-.8pt}{\makebox(0,0){$\bullet$}}}
\put(803,346){\raisebox{-.8pt}{\makebox(0,0){$\bullet$}}}
\put(810,307){\raisebox{-.8pt}{\makebox(0,0){$\bullet$}}}
\put(817,257){\raisebox{-.8pt}{\makebox(0,0){$\bullet$}}}
\put(824,307){\raisebox{-.8pt}{\makebox(0,0){$\bullet$}}}
\put(831,320){\raisebox{-.8pt}{\makebox(0,0){$\bullet$}}}
\put(837,288){\raisebox{-.8pt}{\makebox(0,0){$\bullet$}}}
\put(844,346){\raisebox{-.8pt}{\makebox(0,0){$\bullet$}}}
\put(851,307){\raisebox{-.8pt}{\makebox(0,0){$\bullet$}}}
\put(858,257){\raisebox{-.8pt}{\makebox(0,0){$\bullet$}}}
\put(864,307){\raisebox{-.8pt}{\makebox(0,0){$\bullet$}}}
\put(871,288){\raisebox{-.8pt}{\makebox(0,0){$\bullet$}}}
\put(878,288){\raisebox{-.8pt}{\makebox(0,0){$\bullet$}}}
\put(891,257){\raisebox{-.8pt}{\makebox(0,0){$\bullet$}}}
\put(898,320){\raisebox{-.8pt}{\makebox(0,0){$\bullet$}}}
\put(919,257){\raisebox{-.8pt}{\makebox(0,0){$\bullet$}}}
\put(925,288){\raisebox{-.8pt}{\makebox(0,0){$\bullet$}}}
\put(939,288){\raisebox{-.8pt}{\makebox(0,0){$\bullet$}}}
\put(946,307){\raisebox{-.8pt}{\makebox(0,0){$\bullet$}}}
\put(959,257){\raisebox{-.8pt}{\makebox(0,0){$\bullet$}}}
\put(973,257){\raisebox{-.8pt}{\makebox(0,0){$\bullet$}}}
\put(979,257){\raisebox{-.8pt}{\makebox(0,0){$\bullet$}}}
\put(993,257){\raisebox{-.8pt}{\makebox(0,0){$\bullet$}}}
\put(1007,257){\raisebox{-.8pt}{\makebox(0,0){$\bullet$}}}
\put(1034,257){\raisebox{-.8pt}{\makebox(0,0){$\bullet$}}}
\put(1047,257){\raisebox{-.8pt}{\makebox(0,0){$\bullet$}}}
\put(1061,257){\raisebox{-.8pt}{\makebox(0,0){$\bullet$}}}
\put(1074,257){\raisebox{-.8pt}{\makebox(0,0){$\bullet$}}}
\put(1095,257){\raisebox{-.8pt}{\makebox(0,0){$\bullet$}}}
\put(1115,257){\raisebox{-.8pt}{\makebox(0,0){$\bullet$}}}
\put(1332,257){\raisebox{-.8pt}{\makebox(0,0){$\bullet$}}}
\put(228,815){\usebox{\plotpoint}}
\multiput(228.58,804.02)(0.492,-3.279){19}{\rule{0.118pt}{2.645pt}}
\multiput(227.17,809.51)(11.000,-64.509){2}{\rule{0.400pt}{1.323pt}}
\multiput(239.58,739.00)(0.492,-1.722){19}{\rule{0.118pt}{1.445pt}}
\multiput(238.17,742.00)(11.000,-34.000){2}{\rule{0.400pt}{0.723pt}}
\multiput(250.58,703.81)(0.492,-1.156){19}{\rule{0.118pt}{1.009pt}}
\multiput(249.17,705.91)(11.000,-22.906){2}{\rule{0.400pt}{0.505pt}}
\multiput(261.58,679.57)(0.492,-0.920){19}{\rule{0.118pt}{0.827pt}}
\multiput(260.17,681.28)(11.000,-18.283){2}{\rule{0.400pt}{0.414pt}}
\multiput(272.58,660.23)(0.492,-0.712){21}{\rule{0.119pt}{0.667pt}}
\multiput(271.17,661.62)(12.000,-15.616){2}{\rule{0.400pt}{0.333pt}}
\multiput(284.58,643.32)(0.492,-0.684){19}{\rule{0.118pt}{0.645pt}}
\multiput(283.17,644.66)(11.000,-13.660){2}{\rule{0.400pt}{0.323pt}}
\multiput(295.58,628.62)(0.492,-0.590){19}{\rule{0.118pt}{0.573pt}}
\multiput(294.17,629.81)(11.000,-11.811){2}{\rule{0.400pt}{0.286pt}}
\multiput(306.58,615.77)(0.492,-0.543){19}{\rule{0.118pt}{0.536pt}}
\multiput(305.17,616.89)(11.000,-10.887){2}{\rule{0.400pt}{0.268pt}}
\multiput(317.00,604.92)(0.496,-0.492){19}{\rule{0.500pt}{0.118pt}}
\multiput(317.00,605.17)(9.962,-11.000){2}{\rule{0.250pt}{0.400pt}}
\multiput(328.00,593.92)(0.547,-0.491){17}{\rule{0.540pt}{0.118pt}}
\multiput(328.00,594.17)(9.879,-10.000){2}{\rule{0.270pt}{0.400pt}}
\multiput(339.00,583.92)(0.547,-0.491){17}{\rule{0.540pt}{0.118pt}}
\multiput(339.00,584.17)(9.879,-10.000){2}{\rule{0.270pt}{0.400pt}}
\multiput(350.00,573.93)(0.669,-0.489){15}{\rule{0.633pt}{0.118pt}}
\multiput(350.00,574.17)(10.685,-9.000){2}{\rule{0.317pt}{0.400pt}}
\multiput(362.00,564.93)(0.692,-0.488){13}{\rule{0.650pt}{0.117pt}}
\multiput(362.00,565.17)(9.651,-8.000){2}{\rule{0.325pt}{0.400pt}}
\multiput(373.00,556.93)(0.611,-0.489){15}{\rule{0.589pt}{0.118pt}}
\multiput(373.00,557.17)(9.778,-9.000){2}{\rule{0.294pt}{0.400pt}}
\multiput(384.00,547.93)(0.798,-0.485){11}{\rule{0.729pt}{0.117pt}}
\multiput(384.00,548.17)(9.488,-7.000){2}{\rule{0.364pt}{0.400pt}}
\multiput(395.00,540.93)(0.692,-0.488){13}{\rule{0.650pt}{0.117pt}}
\multiput(395.00,541.17)(9.651,-8.000){2}{\rule{0.325pt}{0.400pt}}
\multiput(406.00,532.93)(0.798,-0.485){11}{\rule{0.729pt}{0.117pt}}
\multiput(406.00,533.17)(9.488,-7.000){2}{\rule{0.364pt}{0.400pt}}
\multiput(417.00,525.93)(0.692,-0.488){13}{\rule{0.650pt}{0.117pt}}
\multiput(417.00,526.17)(9.651,-8.000){2}{\rule{0.325pt}{0.400pt}}
\multiput(428.00,517.93)(0.874,-0.485){11}{\rule{0.786pt}{0.117pt}}
\multiput(428.00,518.17)(10.369,-7.000){2}{\rule{0.393pt}{0.400pt}}
\multiput(440.00,510.93)(0.943,-0.482){9}{\rule{0.833pt}{0.116pt}}
\multiput(440.00,511.17)(9.270,-6.000){2}{\rule{0.417pt}{0.400pt}}
\multiput(451.00,504.93)(0.798,-0.485){11}{\rule{0.729pt}{0.117pt}}
\multiput(451.00,505.17)(9.488,-7.000){2}{\rule{0.364pt}{0.400pt}}
\multiput(462.00,497.93)(0.798,-0.485){11}{\rule{0.729pt}{0.117pt}}
\multiput(462.00,498.17)(9.488,-7.000){2}{\rule{0.364pt}{0.400pt}}
\multiput(473.00,490.93)(0.943,-0.482){9}{\rule{0.833pt}{0.116pt}}
\multiput(473.00,491.17)(9.270,-6.000){2}{\rule{0.417pt}{0.400pt}}
\multiput(484.00,484.93)(0.943,-0.482){9}{\rule{0.833pt}{0.116pt}}
\multiput(484.00,485.17)(9.270,-6.000){2}{\rule{0.417pt}{0.400pt}}
\multiput(495.00,478.93)(1.033,-0.482){9}{\rule{0.900pt}{0.116pt}}
\multiput(495.00,479.17)(10.132,-6.000){2}{\rule{0.450pt}{0.400pt}}
\multiput(507.00,472.93)(0.943,-0.482){9}{\rule{0.833pt}{0.116pt}}
\multiput(507.00,473.17)(9.270,-6.000){2}{\rule{0.417pt}{0.400pt}}
\multiput(518.00,466.93)(0.943,-0.482){9}{\rule{0.833pt}{0.116pt}}
\multiput(518.00,467.17)(9.270,-6.000){2}{\rule{0.417pt}{0.400pt}}
\multiput(529.00,460.93)(0.943,-0.482){9}{\rule{0.833pt}{0.116pt}}
\multiput(529.00,461.17)(9.270,-6.000){2}{\rule{0.417pt}{0.400pt}}
\multiput(540.00,454.93)(0.943,-0.482){9}{\rule{0.833pt}{0.116pt}}
\multiput(540.00,455.17)(9.270,-6.000){2}{\rule{0.417pt}{0.400pt}}
\multiput(551.00,448.93)(1.155,-0.477){7}{\rule{0.980pt}{0.115pt}}
\multiput(551.00,449.17)(8.966,-5.000){2}{\rule{0.490pt}{0.400pt}}
\multiput(562.00,443.93)(0.943,-0.482){9}{\rule{0.833pt}{0.116pt}}
\multiput(562.00,444.17)(9.270,-6.000){2}{\rule{0.417pt}{0.400pt}}
\multiput(573.00,437.93)(1.033,-0.482){9}{\rule{0.900pt}{0.116pt}}
\multiput(573.00,438.17)(10.132,-6.000){2}{\rule{0.450pt}{0.400pt}}
\multiput(585.00,431.93)(1.155,-0.477){7}{\rule{0.980pt}{0.115pt}}
\multiput(585.00,432.17)(8.966,-5.000){2}{\rule{0.490pt}{0.400pt}}
\multiput(596.00,426.93)(1.155,-0.477){7}{\rule{0.980pt}{0.115pt}}
\multiput(596.00,427.17)(8.966,-5.000){2}{\rule{0.490pt}{0.400pt}}
\multiput(607.00,421.93)(0.943,-0.482){9}{\rule{0.833pt}{0.116pt}}
\multiput(607.00,422.17)(9.270,-6.000){2}{\rule{0.417pt}{0.400pt}}
\multiput(618.00,415.93)(1.155,-0.477){7}{\rule{0.980pt}{0.115pt}}
\multiput(618.00,416.17)(8.966,-5.000){2}{\rule{0.490pt}{0.400pt}}
\multiput(629.00,410.93)(1.155,-0.477){7}{\rule{0.980pt}{0.115pt}}
\multiput(629.00,411.17)(8.966,-5.000){2}{\rule{0.490pt}{0.400pt}}
\multiput(640.00,405.93)(1.155,-0.477){7}{\rule{0.980pt}{0.115pt}}
\multiput(640.00,406.17)(8.966,-5.000){2}{\rule{0.490pt}{0.400pt}}
\multiput(651.00,400.93)(1.033,-0.482){9}{\rule{0.900pt}{0.116pt}}
\multiput(651.00,401.17)(10.132,-6.000){2}{\rule{0.450pt}{0.400pt}}
\multiput(663.00,394.93)(1.155,-0.477){7}{\rule{0.980pt}{0.115pt}}
\multiput(663.00,395.17)(8.966,-5.000){2}{\rule{0.490pt}{0.400pt}}
\multiput(674.00,389.93)(1.155,-0.477){7}{\rule{0.980pt}{0.115pt}}
\multiput(674.00,390.17)(8.966,-5.000){2}{\rule{0.490pt}{0.400pt}}
\multiput(685.00,384.93)(1.155,-0.477){7}{\rule{0.980pt}{0.115pt}}
\multiput(685.00,385.17)(8.966,-5.000){2}{\rule{0.490pt}{0.400pt}}
\multiput(696.00,379.93)(1.155,-0.477){7}{\rule{0.980pt}{0.115pt}}
\multiput(696.00,380.17)(8.966,-5.000){2}{\rule{0.490pt}{0.400pt}}
\multiput(707.00,374.93)(1.155,-0.477){7}{\rule{0.980pt}{0.115pt}}
\multiput(707.00,375.17)(8.966,-5.000){2}{\rule{0.490pt}{0.400pt}}
\multiput(718.00,369.93)(1.267,-0.477){7}{\rule{1.060pt}{0.115pt}}
\multiput(718.00,370.17)(9.800,-5.000){2}{\rule{0.530pt}{0.400pt}}
\multiput(730.00,364.94)(1.505,-0.468){5}{\rule{1.200pt}{0.113pt}}
\multiput(730.00,365.17)(8.509,-4.000){2}{\rule{0.600pt}{0.400pt}}
\multiput(741.00,360.93)(1.155,-0.477){7}{\rule{0.980pt}{0.115pt}}
\multiput(741.00,361.17)(8.966,-5.000){2}{\rule{0.490pt}{0.400pt}}
\multiput(752.00,355.93)(1.155,-0.477){7}{\rule{0.980pt}{0.115pt}}
\multiput(752.00,356.17)(8.966,-5.000){2}{\rule{0.490pt}{0.400pt}}
\multiput(763.00,350.93)(1.155,-0.477){7}{\rule{0.980pt}{0.115pt}}
\multiput(763.00,351.17)(8.966,-5.000){2}{\rule{0.490pt}{0.400pt}}
\multiput(774.00,345.93)(1.155,-0.477){7}{\rule{0.980pt}{0.115pt}}
\multiput(774.00,346.17)(8.966,-5.000){2}{\rule{0.490pt}{0.400pt}}
\multiput(785.00,340.94)(1.505,-0.468){5}{\rule{1.200pt}{0.113pt}}
\multiput(785.00,341.17)(8.509,-4.000){2}{\rule{0.600pt}{0.400pt}}
\multiput(796.00,336.93)(1.267,-0.477){7}{\rule{1.060pt}{0.115pt}}
\multiput(796.00,337.17)(9.800,-5.000){2}{\rule{0.530pt}{0.400pt}}
\multiput(808.00,331.93)(1.155,-0.477){7}{\rule{0.980pt}{0.115pt}}
\multiput(808.00,332.17)(8.966,-5.000){2}{\rule{0.490pt}{0.400pt}}
\multiput(819.00,326.94)(1.505,-0.468){5}{\rule{1.200pt}{0.113pt}}
\multiput(819.00,327.17)(8.509,-4.000){2}{\rule{0.600pt}{0.400pt}}
\multiput(830.00,322.93)(1.155,-0.477){7}{\rule{0.980pt}{0.115pt}}
\multiput(830.00,323.17)(8.966,-5.000){2}{\rule{0.490pt}{0.400pt}}
\multiput(841.00,317.93)(1.155,-0.477){7}{\rule{0.980pt}{0.115pt}}
\multiput(841.00,318.17)(8.966,-5.000){2}{\rule{0.490pt}{0.400pt}}
\multiput(852.00,312.94)(1.505,-0.468){5}{\rule{1.200pt}{0.113pt}}
\multiput(852.00,313.17)(8.509,-4.000){2}{\rule{0.600pt}{0.400pt}}
\multiput(863.00,308.93)(1.155,-0.477){7}{\rule{0.980pt}{0.115pt}}
\multiput(863.00,309.17)(8.966,-5.000){2}{\rule{0.490pt}{0.400pt}}
\multiput(874.00,303.94)(1.651,-0.468){5}{\rule{1.300pt}{0.113pt}}
\multiput(874.00,304.17)(9.302,-4.000){2}{\rule{0.650pt}{0.400pt}}
\multiput(886.00,299.93)(1.155,-0.477){7}{\rule{0.980pt}{0.115pt}}
\multiput(886.00,300.17)(8.966,-5.000){2}{\rule{0.490pt}{0.400pt}}
\multiput(897.00,294.94)(1.505,-0.468){5}{\rule{1.200pt}{0.113pt}}
\multiput(897.00,295.17)(8.509,-4.000){2}{\rule{0.600pt}{0.400pt}}
\multiput(908.00,290.93)(1.155,-0.477){7}{\rule{0.980pt}{0.115pt}}
\multiput(908.00,291.17)(8.966,-5.000){2}{\rule{0.490pt}{0.400pt}}
\multiput(919.00,285.94)(1.505,-0.468){5}{\rule{1.200pt}{0.113pt}}
\multiput(919.00,286.17)(8.509,-4.000){2}{\rule{0.600pt}{0.400pt}}
\multiput(930.00,281.93)(1.155,-0.477){7}{\rule{0.980pt}{0.115pt}}
\multiput(930.00,282.17)(8.966,-5.000){2}{\rule{0.490pt}{0.400pt}}
\multiput(941.00,276.94)(1.651,-0.468){5}{\rule{1.300pt}{0.113pt}}
\multiput(941.00,277.17)(9.302,-4.000){2}{\rule{0.650pt}{0.400pt}}
\multiput(953.00,272.93)(1.155,-0.477){7}{\rule{0.980pt}{0.115pt}}
\multiput(953.00,273.17)(8.966,-5.000){2}{\rule{0.490pt}{0.400pt}}
\multiput(964.00,267.94)(1.505,-0.468){5}{\rule{1.200pt}{0.113pt}}
\multiput(964.00,268.17)(8.509,-4.000){2}{\rule{0.600pt}{0.400pt}}
\multiput(975.00,263.93)(1.155,-0.477){7}{\rule{0.980pt}{0.115pt}}
\multiput(975.00,264.17)(8.966,-5.000){2}{\rule{0.490pt}{0.400pt}}
\multiput(986.00,258.94)(1.505,-0.468){5}{\rule{1.200pt}{0.113pt}}
\multiput(986.00,259.17)(8.509,-4.000){2}{\rule{0.600pt}{0.400pt}}
\multiput(997.00,254.94)(1.505,-0.468){5}{\rule{1.200pt}{0.113pt}}
\multiput(997.00,255.17)(8.509,-4.000){2}{\rule{0.600pt}{0.400pt}}
\multiput(1008.00,250.93)(1.155,-0.477){7}{\rule{0.980pt}{0.115pt}}
\multiput(1008.00,251.17)(8.966,-5.000){2}{\rule{0.490pt}{0.400pt}}
\multiput(1019.00,245.94)(1.651,-0.468){5}{\rule{1.300pt}{0.113pt}}
\multiput(1019.00,246.17)(9.302,-4.000){2}{\rule{0.650pt}{0.400pt}}
\multiput(1031.00,241.94)(1.505,-0.468){5}{\rule{1.200pt}{0.113pt}}
\multiput(1031.00,242.17)(8.509,-4.000){2}{\rule{0.600pt}{0.400pt}}
\multiput(1042.00,237.93)(1.155,-0.477){7}{\rule{0.980pt}{0.115pt}}
\multiput(1042.00,238.17)(8.966,-5.000){2}{\rule{0.490pt}{0.400pt}}
\multiput(1053.00,232.94)(1.505,-0.468){5}{\rule{1.200pt}{0.113pt}}
\multiput(1053.00,233.17)(8.509,-4.000){2}{\rule{0.600pt}{0.400pt}}
\multiput(1064.00,228.94)(1.505,-0.468){5}{\rule{1.200pt}{0.113pt}}
\multiput(1064.00,229.17)(8.509,-4.000){2}{\rule{0.600pt}{0.400pt}}
\multiput(1075.00,224.94)(1.505,-0.468){5}{\rule{1.200pt}{0.113pt}}
\multiput(1075.00,225.17)(8.509,-4.000){2}{\rule{0.600pt}{0.400pt}}
\multiput(1086.00,220.93)(1.155,-0.477){7}{\rule{0.980pt}{0.115pt}}
\multiput(1086.00,221.17)(8.966,-5.000){2}{\rule{0.490pt}{0.400pt}}
\multiput(1097.00,215.94)(1.651,-0.468){5}{\rule{1.300pt}{0.113pt}}
\multiput(1097.00,216.17)(9.302,-4.000){2}{\rule{0.650pt}{0.400pt}}
\multiput(1109.00,211.94)(1.505,-0.468){5}{\rule{1.200pt}{0.113pt}}
\multiput(1109.00,212.17)(8.509,-4.000){2}{\rule{0.600pt}{0.400pt}}
\multiput(1120.00,207.94)(1.505,-0.468){5}{\rule{1.200pt}{0.113pt}}
\multiput(1120.00,208.17)(8.509,-4.000){2}{\rule{0.600pt}{0.400pt}}
\multiput(1131.00,203.93)(1.155,-0.477){7}{\rule{0.980pt}{0.115pt}}
\multiput(1131.00,204.17)(8.966,-5.000){2}{\rule{0.490pt}{0.400pt}}
\multiput(1142.00,198.94)(1.505,-0.468){5}{\rule{1.200pt}{0.113pt}}
\multiput(1142.00,199.17)(8.509,-4.000){2}{\rule{0.600pt}{0.400pt}}
\multiput(1153.00,194.94)(1.505,-0.468){5}{\rule{1.200pt}{0.113pt}}
\multiput(1153.00,195.17)(8.509,-4.000){2}{\rule{0.600pt}{0.400pt}}
\multiput(1164.00,190.94)(1.651,-0.468){5}{\rule{1.300pt}{0.113pt}}
\multiput(1164.00,191.17)(9.302,-4.000){2}{\rule{0.650pt}{0.400pt}}
\multiput(1176.00,186.93)(1.155,-0.477){7}{\rule{0.980pt}{0.115pt}}
\multiput(1176.00,187.17)(8.966,-5.000){2}{\rule{0.490pt}{0.400pt}}
\multiput(1187.00,181.94)(1.505,-0.468){5}{\rule{1.200pt}{0.113pt}}
\multiput(1187.00,182.17)(8.509,-4.000){2}{\rule{0.600pt}{0.400pt}}
\multiput(1198.00,177.94)(1.505,-0.468){5}{\rule{1.200pt}{0.113pt}}
\multiput(1198.00,178.17)(8.509,-4.000){2}{\rule{0.600pt}{0.400pt}}
\multiput(1209.00,173.94)(1.505,-0.468){5}{\rule{1.200pt}{0.113pt}}
\multiput(1209.00,174.17)(8.509,-4.000){2}{\rule{0.600pt}{0.400pt}}
\multiput(1220.00,169.94)(1.505,-0.468){5}{\rule{1.200pt}{0.113pt}}
\multiput(1220.00,170.17)(8.509,-4.000){2}{\rule{0.600pt}{0.400pt}}
\multiput(1231.00,165.94)(1.505,-0.468){5}{\rule{1.200pt}{0.113pt}}
\multiput(1231.00,166.17)(8.509,-4.000){2}{\rule{0.600pt}{0.400pt}}
\multiput(1242.00,161.94)(1.651,-0.468){5}{\rule{1.300pt}{0.113pt}}
\multiput(1242.00,162.17)(9.302,-4.000){2}{\rule{0.650pt}{0.400pt}}
\multiput(1254.00,157.93)(1.155,-0.477){7}{\rule{0.980pt}{0.115pt}}
\multiput(1254.00,158.17)(8.966,-5.000){2}{\rule{0.490pt}{0.400pt}}
\multiput(1265.00,152.94)(1.505,-0.468){5}{\rule{1.200pt}{0.113pt}}
\multiput(1265.00,153.17)(8.509,-4.000){2}{\rule{0.600pt}{0.400pt}}
\multiput(1276.00,148.94)(1.505,-0.468){5}{\rule{1.200pt}{0.113pt}}
\multiput(1276.00,149.17)(8.509,-4.000){2}{\rule{0.600pt}{0.400pt}}
\multiput(1287.00,144.94)(1.505,-0.468){5}{\rule{1.200pt}{0.113pt}}
\multiput(1287.00,145.17)(8.509,-4.000){2}{\rule{0.600pt}{0.400pt}}
\multiput(1298.00,140.94)(1.505,-0.468){5}{\rule{1.200pt}{0.113pt}}
\multiput(1298.00,141.17)(8.509,-4.000){2}{\rule{0.600pt}{0.400pt}}
\multiput(1309.00,136.94)(1.505,-0.468){5}{\rule{1.200pt}{0.113pt}}
\multiput(1309.00,137.17)(8.509,-4.000){2}{\rule{0.600pt}{0.400pt}}
\multiput(1320.00,132.94)(1.651,-0.468){5}{\rule{1.300pt}{0.113pt}}
\multiput(1320.00,133.17)(9.302,-4.000){2}{\rule{0.650pt}{0.400pt}}
\end{picture}

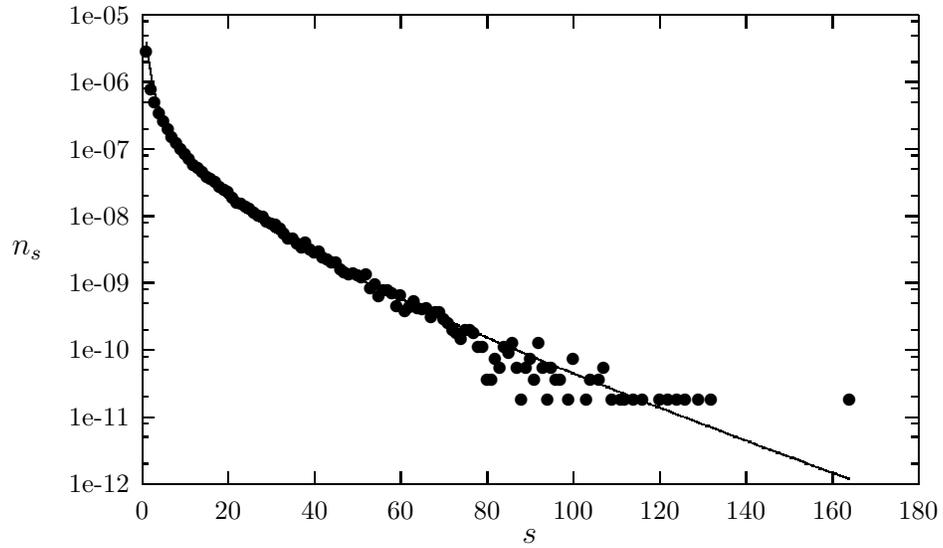
\noindent {\bf Fig.1.}  The normalised distribution ($n_s$) of clusters of size $s$
for the automaton value $k=3$. Solid line represents the best fit.
\end{document}